\documentclass[12pt]{iopart}
\usepackage{iopams}
\usepackage{graphicx}
\gdef\ffrac#1#2{\textstyle{#1\over#2}\displaystyle}
\gdef\no#1{:\!#1\!:}

\begin{document}
\title[Quantum Quenches to a Critical Point in One Dimension]
{Quantum Quenches to a Critical Point in One Dimension: some further results}
\author{John Cardy$^{1,2}$}
\address{$^1$Rudolf
Peierls Centre for Theoretical Physics, 1 Keble Road, Oxford OX1
3NP, UK}
\address{$^2$All Souls College, Oxford}
\ead{j.cardy1@physics.ox.ac.uk}

\begin{abstract}
We describe several results concerning global quantum quenches from states with short-range correlations to quantum critical points whose low-energy properties are described by a 1+1-dimensional conformal field theory (CFT), extending the work of Calabrese and Cardy (2006): (a) for the special class of initial states discussed in that paper we show that, once a finite region falls inside the horizon, its reduced density matrix is exponentially close in $L_2$ norm to that of a thermal Gibbs state; (b) small deformations of this initial state in general lead to a (non-Abelian) generalized Gibbs distribution (GGE) with, however, the possibility of parafermionic conserved charges; (c) small deformations of the CFT, corresponding to curvature of the dispersion relation and (non-integrable) left-right scattering, lead to a dependence of the speed of propagation on the initial state, as well as diffusive broadening of the horizon.

\end{abstract}

\maketitle

\section{Introduction}

In a global quantum quench, a quantum system extended in space is initially prepared in a pure state $|\psi_0\rangle$, which is usually taken to be the ground state of hamiltonian $H_0$ with short-range interactions. It is then evolved unitarily with a hamiltonian $H$, which usually differs from $H_0$ by the quench, or sudden change, of some parameter. Such a protocol is an idealization of real experiments which may be carried out, for example, on ultra-cold atoms. However, for the purposes of this paper, the important features of $|\psi_0\rangle$ are that it is translationally invariant in the thermodynamic limit and that it has short-ranged correlations and area-law entanglement entropy. Furthermore we consider only the case when $H$ is gapless, corresponding to a quantum critical point, and with a linear low-energy dispersion relation $\omega\sim v|k|$. In that case the low-energy spectrum is described by a conformal field theory (CFT) with hamiltonian $H_{CFT}$. 

A great deal is known about the properties of CFTs, especially in 1+1 dimensions. In 2006  Calabrese and this author \cite{CC} (CC) used this knowledge to derive properties of a quench to a quantum critical point whose entire spectrum is described exactly by a CFT, that is assuming that $H\equiv H_{CFT}$. They also made a particular ansatz for the initial state $|\psi_0\rangle$.  
Their main results were: 
\begin{itemize}
\item 1-point functions of non-conserved local observables in general decay exponentially to their values in the ground state of $H$; 
\item the `horizon effect': two- and higher point functions of local observables whose arguments lie in a finite interval $A$ of length $\ell$ become stationary after a time $t\sim\ell/2v$ when the left- and rightmost points first fall into a forward light cone whose apex lies on $t=0$; 
\item thermalization: after that time they are equal, up to exponentially small corrections, to the values they would have at the finite temperature $\beta^{-1}$ in the CFT corresponding to the mean energy $E_0=\langle\psi_0|H_{CFT}|\psi_0\rangle$. This is equivalent to the statement that reduced density matrix of the interval $\rho_\ell\propto {\rm Tr}_{\overline A}\,e^{-\beta H_{CFT}}$; 
\item although the whole system remains in a pure state, the entanglement entropy of the interval with the rest of the system increases linearly with $t$ up to the equilibration time, after which it becomes stationary and $\propto\ell$: the extensive part is equal to the thermal Gibbs entropy at temperature $\beta^{-1}$ \cite{CC2}.
\item the horizon itself has width $O(\beta)$, and it is only inside this relatively narrow region that the details of the particular CFT become relevant.
\end{itemize}

Although these results depend on the specific assumptions discussed above, as well as detailed analyticity properties of the CFT in complex space-time, they may be physically interpreted rather simply in terms of the emission of left- and right-moving pairs of quasiparticles from the initial state, which are correlated and entangled over only lengths $O(\beta)$, and which otherwise move semi-classically. It turns out that, suitably adapted to account for the dispersion relation, this quasiparticle picture extends to other hamiltonians, including lattice models and to non-critical quenches, at least as long as the hamiltonian $H$ is integrable so that even the high-energy states have a quasiparticle description. The horizon effect is supposed to hold more generally in most systems as a consequence of the Lieb-Robinson bound \cite{LR}, and has been observed qualitatively in experiments \cite{expt}. The general picture of thermalization has also been discussed within the AdS/CFT correspondence, when it is supposed to be a consequence of the formation of a black hole \cite{AdS}.

However this leaves several unanswered questions, some of which we address in this paper:
\begin{enumerate}
\item can we quantify how close $\rho_\ell$ is to a thermal density matrix?
\item in an integrable model, there are an infinite number of conserved quantities which commute with the hamiltonian $H$. In that case, it has been suggested that stationary state should be described by a generalized Gibbs ensemble (GGE) rather than a simple Gibbs distribution \cite{GGE}. This has been the subject of intense investigation in several solvable models \cite{solvable}. How does this appear within a CFT, which (at least in the rational case) is the most integrable model of all?
\item what happens if we relax the particular ansatz that CC made for the initial state?
\item what happens if we add irrelevant interactions to $H_{CFT}$ which give only non-leading corrections to the low-energy behaviour? How do they affect the quench dynamics?
\item what happens in a finite system? Is it possible for the initial state to fully or partially revive?
\end{enumerate}

In what follows we shall show that, under the assumptions of CC, the answer to (i) is that the overlap between $\rho_\ell$ and the reduced density matrix $\tilde\rho_\beta$ in a thermal ensemble is exponentially close to unity, that is
\begin{equation}\label{close}
1-\frac{{\rm Tr}\,(\rho_\ell\cdot\tilde\rho_\beta)}{\big(({\rm Tr}\,\rho_\ell^2)({\rm Tr}\,\tilde\rho_\beta^2)\big)^{1/2}}
<{\rm const.}\,e^{-2\pi\Delta(t-\ell/2v)/\beta}\,.
\end{equation}
where $\Delta$ is a universal exponent depending on the initial state. (The left hand side is of course always $\geq0$ by the Schwarz inequality.)

The resolution of (ii) and (iii) is as follows: the ansatz of CC amounts to the assumption 
$$
|\psi_0\rangle\propto e^{-(\beta/4)H_{CFT}}|B\rangle\,,
$$
where $|B\rangle$ is a conformal boundary state. (The factor $\beta/4$ is chosen so that the expectation of $H_{CFT}$ in this state is the same as that at finite temperature $\beta^{-1}$.) We remind the reader that each bulk CFT is supposed to have a particular allowed set of such boundary states, each of which corresponds to a fixed point of the boundary renormalization group (RG),  each with their own basins of attraction. For the minimal models, there is a finite number of these states, and their basins of attraction are supposed to contain the ground states of each possible non-critical hamiltonian $H_0$. Thus this ground state should be representable in terms of all possible irrelevant operators acting on $|B\rangle$:
\begin{equation}\label{generalstate}
|\psi_0\rangle\propto \prod_ke^{-\beta_k\int\widetilde\Phi_k dx}|B\rangle\,,
\end{equation}
suitably regularized. One of these irrelevant operators is always the component $T_{tt}$ of the stress tensor, whose space integral is the hamiltonian $H_{CFT}$. Thus the CC ansatz is equivalent to assuming that this is the \em only \em term in the above sum. $T_{tt}$ may be written as a sum of holomorphic and antiholomorphic operators $T$ and $\overline T$, each of which is conserved and whose space integral is a conserved charge. This is equally true of all the other irrelevant operators which are its descendants, which can be written as powers of $T$ and derivatives thereof, plus their antiholomorphic partners. We shall argue that in these cases the quench dynamics from such a state with a hamiltonian $H=H_{CFT}$ leads to a stationary reduced density matrix for a finite interval which has the GGE form
\begin{equation}\label{GGE}
\rho_\ell\propto {\rm Tr}_{\overline A}\,\prod_ke^{-\beta_k\cos(\pi\Delta_k/2)\int(\Phi_k+\overline\Phi_k) dx}\,,
\end{equation}
where  now $\Phi_k$ and $\overline\Phi_k$ are  holomorphic and antiholomorphic \em bulk \em operators. so that their space integrals are conserved charges. Note, however, that the charges $\int(\Phi_k+\overline\Phi_k) dx$ above do not necessarily commute among themselves, even though they commute with the hamiltonian $H_{CFT}$. The conventional GGE includes only a commuting sub-algebra of these charges. This is motivated by the idea that these should form a complete set of macroscopic commuting observables which should characterize any macrostate. However, this does not appear to be the case for a 1+1-dimensional CFT: because of the exactly linear dispersion relation, there is a massive degeneracy of states, and the expectation values of the charges in the commuting sub-algebra (which were identified in \cite{Zam}) are not sufficient to characterize the states.  A similar phenomenon has been pointed out by Sotiriadis \cite{Sot} in the context of a quench in a 1+1-dimensional massless free boson theory from a non-gaussian initial state, although in this case the commuting conserved charges are the mode occupation numbers which are non-local.

However, this still does not account for all possible irrelevant boundary operators in (\ref{generalstate}). We further argue that, at least in rational CFTs, for every boundary operator $\widetilde\Phi_k$ there is a pair of holomorphic and antiholomorphic operators $(\Phi_k,\overline\Phi_k)$ with the same overall scaling dimension $\Delta_k$. These also lead to conserved charges and a generalization of (\ref{GGE}).
However, since $\Delta_k$ is in general non-integer, these currents are parafermionic and the corresponding charges are non-local, acting, for example, to change the boundary conditions. Although such charges are not customarily included in the GGE, we shall argue that they make perfect sense, at least perturbatively in the $\beta_k$, and that they lead to physically sensible results.  

Question (iv) seems to be difficult to address in general. A perturbative approach would require knowing the correlators of all possible irrelevant operators after the quench. However, in the case of operators $(T^2,\overline T^2,T\overline T)$ involving the square of the stress tensor, which are present in all critical theories with a UV regulator, we can express the effect in terms of a coupling to a random metric with short-range correlations. These terms may be thought of as introducing soft scattering of the original CFT quasiparticles. The first two terms break the low-energy Lorentz invariance and correspond to curvature of the dispersion relation. The $T\overline T$ term corresponds to left-right scattering and is the simplest non-integrable perturbation of the CFT. The geometrical interpretation of these terms leads to two main effects: a modification of the propagation velocity $v$ dependent on the effective temperature $\beta^{-1}$ of the initial state, and a spreading of the horizon width as $(t\beta)^{1/2}$. Although strictly valid only at weak coupling, it may be argued that this general picture should persist to higher orders and for other irrelevant operators which are powers and derivatives of the stress tensor.
Dependence of the propagation speed on the initial state has been observed in several studies of lattice models \cite{vdep}.

(v) has previously been studied in Ref.~\cite{JCrevival}. It can be addressed in terms of the return amplitude
$$
{\cal F}(t)=|\langle\psi_0|e^{-iHt}|\psi_0\rangle|\,.
$$
In a periodic system of length $L$ (and also in an open one with conformal boundary conditions) in pure CFT the eigenvalues of $H$ are of the form $(2\pi/L)(\Delta+{\rm integer})$, and therefore in a rational CFT, when the scaling dimensions $\Delta$ take only a finite number of rational values $N/M$, there will always be complete revivals with ${\cal F}=1$ at times such that $2vt=ML$ ($vt$ for an open system). However, whether there are partial revivals with ${\cal F}=O(1)$ at earlier times depends on details of the CFT. For the CC initial state it turns out that ${\cal F}(t)$ is given by the continuation of the partition function of the CFT on an annulus, or rectangle, continued to complex values of the  modulus, or aspect ratio. For rational CFTs this leads to partial revivals whenever $2vt/L$ is an integer, and in fact a remarkably rich structure near every rational value of $vt/L$, related to properties of the CFT characters under the modular group. If a weak perturbation such as $T\overline T$ is added to the hamiltonian the effect is to broaden the revivals with a width $\propto t^{1/2}$.

The layout of this paper is as follows. In Sec.~\ref{sec2} we recall the arguments of CC, in particular rephrasing the basic assumption about the initial state made in those papers, then recall the main results, their physical interpretation, and the technical properties of CFT that allow these arguments to go through.
Once this is set up the statement about the convergence of the reduced density matrix $\rho_\ell$ to the thermal distribution $\tilde\rho_\beta$ follows fairly straightforwardly.

Sec.~\ref{sec3} addresses deviations from the CC initial state, and whether they lead to a GGE. We first restrict to the case of irrelevant operators related to the stress tensor and derivatives and powers thereof. We then study the properties of this GGE in CFT in general, and show, for example,  that it leads to observable-dependent effective temperatures. Since these arguments require the use of possibly delicate analytic continuations within the path integral, it is reassuring that the results can also be obtained by a straightforward resummation of perturbation theory.  This also leads to a simple physical interpretation of the results of this section.

We then discuss the slightly more tricky question of more general perturbations of the CC state with non-integer dimensions. We argue that their effects may only be described within a GGE if semi-local charges which are integrals of parafermionic currents are included. We give an illustrative example of the transverse field Ising model quenched to the critical point from a short-range disordered state with a small non-zero longitudinal magnetization.

Finally in Sec.~\ref{sec4} we consider irrelevant perturbations of the evolving hamiltonian $H_{CFT}$, specifically those second order in the stress tensor. We argue that these are equivalent to coupling to a random metric, and, to lowest order are able to deduce the results described earlier. We then argue that the qualitative picture they give remains valid at higher orders and for more general perturbations.

\section{Convergence to a thermal ensemble.}\label{sec2}
\subsection{Correlation functions.}
In this section we revisit the arguments of Calabrese and Cardy \cite{CC} [CC] , reformulating them in a way that makes the assumptions more transparent, and then argue that the reduced density matrix $\rho_\ell$ of a finite interval,  after falling within the horizon, is exponentially close to that of  a thermal ensemble. 

We consider the time evolution of a one-dimensional quantum system from an initial state $|\psi_0\rangle$, assumed to be translationally invariant, with short-range correlations and entanglement. The system evolves unitarily for times $t>0$ with the hamiltonian $H$, which we take to be that of a 1+1-dimensional CFT. Thus the state at time $t$ in the Schr\"odinger picture is $e^{-itH}|\psi_0\rangle$ with $H=H_{CFT}$. We are primarily interested in correlation functions of local operators $\Phi_j(x_j)$, $1\leq j\leq n$, possibly evaluated at different times, which will therefore be of the form
$$
\langle\psi_0|e^{it_nH}\Phi_n(x_n)e^{-i(t_n-t_{n-1})H}\ldots \Phi_2(x_2)e^{-i(t_2-t_1)H} \Phi_1(x_1) e^{-it_1H} |\psi_0\rangle\,,
$$
where $t_n\geq t_{n-1}\geq\ldots\geq t_1\geq0$.
Since CFT is most easily formulated in euclidean space, it is simpler to start from the imaginary time evolution
$$
\langle\psi_0|e^{-\tau_f H}\Phi_n(x_n)e^{-(\tau_n-\tau_{n-1})H}\ldots \Phi_2(x_2)e^{-(\tau_2-\tau_1)H} \Phi_1(x_1) e^{-\tau_1H} |\psi_0\rangle\,.
$$
This can be considered as a correlation function in the Heisenberg picture
$$
\langle\Phi_n(x_n,\tau_n)\ldots\Phi_2(x_2,\tau_2)\Phi_1(x_1,\tau_1)\rangle\,,
$$
in a euclidean slab geometry $0\leq\tau\leq\tau_{\rm tot}=\tau_n+\tau_f$ with boundary conditions on each edge corresponding to the state $|\psi_0\rangle$. It should then be continued to $\tau_j\to it_j$ and $\tau_f\to-it_n$. The problem with this is that the total width $\tau_{\rm tot}$ of the slab is then zero. In CC this was avoided by appealing to the theory of boundary critical phenomena: the actual boundary conditions at $\tau=0$ and $\tau_{\rm tot}$ are replaced by `idealized' boundary conditions at $\tau=-\tau_0,\tau_{\rm tot}+\tau_0$, where $\tau_0$ is the so-called extrapolation length. In the theory of boundary critical behaviour, this is justified on the basis of the renormalization group (RG): the idealized boundary condition corresponds to a fixed point of the RG, and $\tau_0$ measures the deviation of the actual state from this. For the example of a transverse Ising model initially in the disordered phase, this ideal state would be an unentangled  product state. Once we take a finite $\tau_0$, it then makes sense to take $\tau_{\rm tot}\to0$. We must then compute correlation functions in a slab of width $2\tau_0$ with ideal boundary conditions. However this argument is based on the assumption that the long-time behaviour after the quench should be insensitive to the details of the initial state as long as it has only short-range correlations. As we shall argue, at least for the case of evolution with a CFT hamiltonian, this is not in fact the case.

Let us therefore rephrase the prescription of CC in a way that its assumptions are more clear and so may be generalized. It is equivalent to assuming that the initial state
has the form
\begin{equation}
\label{prescrip}
|\psi_0\rangle\propto e^{-\tau_0H}|B\rangle\,,
\end{equation}
where $|B\rangle$ is the (conformally invariant) state corresponding to the idealized boundary condition, corresponding to a fixed point of the boundary RG. In CFT, such fixed points correspond to conformally invariant boundary conditions, for which the component $T_{\tau x}\propto T-\overline T$ vanishes
$$
\big(T(x)-\overline T(x)\big)|B\rangle=0\,.
$$
For such boundary conditions, correlation functions in the slab are simply related to those in any other simply connected region obtained by a conformal mapping,  in particular the upper half plane $\mathbb H$. 

Note that the expectation value of $H$ in this state is given by the free energy per unit width of a euclidean strip of width $2\tau_0$ \cite{BCN}:
\begin{equation}\label{H}
\langle\psi_0|H|\psi_0\rangle=\frac{\pi cL}{24(2\tau_0)^2}\,,
\end{equation}
where $L$ is the total length, $c$ is the central charge, and the expectation value in the ground state has been normalized to zero. This may be compared with the mean energy at finite temperature $\beta^{-1}$ \cite{BCN,Aff}
$$
\frac{{\rm Tr}\,He^{-\beta H}}{{\rm Tr}\,e^{-\beta H}}=\frac{\pi cL}{6\beta^2}\,.
$$
Since $H$ is conserved  we see, anticipating thermalization, that $\tau_0=\beta/4$ and we adopt this parametrization from now on.

Altough from this point of view, the prescription (\ref{prescrip}) seems rather artificial, as we argued in the introduction it may be viewed as including just one, in fact often the most important, of the possible irrelevant operators in the sum (\ref{generalstate}). For the purposes of the remainder of this chapter we shall assume this to be the case, addressing the other terms in Sec.~\ref{sec3}. In this state correlations decay exponentially (with inverse correlation lengths $\pi\widetilde\Delta/(\beta/2)$ where $\widetilde\Delta$ is a boundary scaling dimension \cite{BCN}) and there is area-law entanglement 
entropy $\propto c\log\beta$. (We assume that $\beta$ is always larger than any microscopic length scale, in units where $v=1$.)

We are therefore faced with the problem of computing correlation functions 
$$
\langle\Phi_n(x_n,\tau_n)\ldots\Phi_2(x_2,\tau_2)\Phi_1(x_1,\tau_1)\rangle
$$
in the strip $-\beta/4<\tau<\beta/4$ with conformal boundary conditions, and analytically continuing them to real time $\tau_j\to it_j$. Fortunately the analytic properties of CFT allow this to be carried out. 

The picture is simpler if we conformally transform the strip: define $w=x+i\tau$ and the conformal mapping
\begin{equation}\label{zw}
w\to z=i\,e^{2\pi w/\beta}\,,
\end{equation}
which sends the interior of the strip to the upper half plane $\mathbb H$. A correlation function of primary operators $\Phi_j$ is given, apart from jacobian factors, by a similar correlation function in $\mathbb H$ which depends on the points $z_j$ and their images $\bar z_j$ in the real axis:
$$
\langle\Phi_n(x_n,\tau_n)\ldots\Phi_2(x_2,\tau_2)\Phi_1(x_1,\tau_1)\rangle\propto F(\{z_j\},\{\bar z_j\})\,.
$$
Moreover, $F(\{z_j\},\{\bar z_j\})$ may be written as a linear combination of conformal blocks 
$\sum_kA_k F_k(\{z_j\},\{\bar z_j\})$, each of which is the analytic continuation to $z'_j=\bar z_j$ of a function $F_k(\{z_j\},\{z'_j\})$ defined on a branched cover of the whole plane ${\mathbb C}^n$. Moreover global conformal invariance constrains $F_k(\{z_j\},\{z'_j\})$ to depend only on the cross-ratios such as
$$
\eta_{ij}=(z_i-z_j)(z_i'-z'_j)/(z_i-z'_i)(z_j-z'_j)\,,
$$
with branch points at coincident points where $\eta_{ij}=0,1,\infty$. (For the two-point function there is only one such cross-ratio.) These blocks can be labelled so that each has a well-defined power-law OPE behaviour as some independent set of the $\eta_{ij}\to0$.\footnote{It has recently been pointed out \cite{Hart} that for higher-order correlation functions in non-rational CFTs the behavior at some of the singular points is not implied by the OPE. This does not affect our later conclusion which uses only the 2-point function of twist operators.}  The leading term in the OPE always corresponds to the identity operator, since this limit corresponds to the case when all the $\{z_j\}$ and the $\{\bar z_j\}$ are very far from the real axis, and each conformal block factorizes into a holomorphic part depending on the $\{z_j\}$ and its complex conjugate. This then gives the bulk correlator of the same product of operators. 

Something remarkable now happens when we continue to real time $\tau_j\to it_j$. In that case
$$
w_j=x_j-t_j\,,\qquad w_j'=x_j+t_j+i\beta/2\,,
$$
and 
$$
z_j=ie^{(2\pi/\beta)(x_j-t_j)}\,,\qquad z'_j=-ie^{(2\pi/\beta)(x_j+t_j)}\,.
$$
Note that these all lie on the imaginary axis in the $z$-plane. However $z'_j$ is now no longer at the image point of $z_j$. 
Suppose now that all the $t_j$ are sufficiently large so that 
$$
(x_j+t_j)-(x_{j'}-t_{j'})\gg\beta, \qquad(\forall(j,j'))\,.
$$
This corresponds to the intersection of all the past light cones with $t=0$ being non-empty, that is all points have fallen inside a horizon, defined as a future light cone originating at some point on $t=0$. In that case it may be checked \cite{CC} that 
$$
|\eta_{jj'}|\ll1, \qquad(\forall(j,j'))\,,
$$
and we may apply the OPE to deduce that in this limit the leading term is the bulk correlation function, that is, the presence of the boundary is irrelevant. If we now reverse the conformal mapping, we see that in this limit when all the points have fallen inside a horizon, the correlator is equal to that on a cylinder of radius $\beta$, that is, the same as at finite temperature. The corrections to this come from higher-order terms in the OPE. In the $z$-plane they decay as $|z|^{-\Delta}$, where $\Delta$ is the bulk scaling dimension of the lowest dimension non-trivial operator to which $\prod_{j=1}^n\Phi_j$ couples, which has a non-zero expectation value near the boundary. This depends on the particular boundary state $|B\rangle$. In the $w$-plane, this corresponds to exponentially decaying corrections of the form
$$
e^{-(\pi\Delta/\beta)\big((x_j+t_j)-(x_{j'}-t_{j'})\big)}\,.
$$

\subsection{Reduced density matrix.}

The above argument leaves open, for example, the question of other correlators, for example those of non-primary operators. A more complete result about thermalization would be that the whole reduced density matrix of an interval is close to that of a thermal ensemble, once the interval has fallen inside the horizon. This would imply that all equal-time correlation functions of local operators with arguments in the interval are close to their thermal values. This result we now derive. 

Consider an interval $(x_1,x_2)$ of length $\ell\ll L$ (in practice it will suffice that $L/2-\ell\gg\beta$.) The reduced density matrix of this interval is in general
$$
\rho_\ell(t)=\frac{{\rm Tr}_{L-\ell}\,e^{-iHt}|\psi_0\rangle\langle\psi_0|e^{iHt}}{{\rm Tr}_{L}\,|\psi_0\rangle\langle\psi_0|}\,,
$$
in an obvious notation where ${\rm Tr}_X$ means a trace over the degrees of freedom in the interval $X$. For the CC state this becomes
$$
\rho_\ell(\tau)=\frac{{\rm Tr}_{L-\ell}\,e^{-H(\beta/4+\tau)}|B\rangle\langle B|e^{-H(\beta/4-\tau)}}{{\rm Tr}_{L}\,e^{-(\beta/4)H}|B\rangle\langle B|e^{-(\beta/4)H}}\,,
$$
continued to $\tau=it$. 

For real $\tau$ the numerator of this expression is the partition function on a strip $\cal S$ slit along $(\ell,\tau)$, the rows and columns of $\rho_\ell$ being labelled by the fixed values of a complete commuting set of local fields above and below the slit. The denominator is the trace ${\rm Tr}_\ell$ of this, equivalent to sewing up the slit. 

Similarly the reduced density matrix in a thermal ensemble 
$$
\tilde\rho_\beta=\frac{{\rm Tr}_{L-\ell}\,e^{-\beta H}}{{\rm Tr}_{L}\,e^{-\beta H}}
$$
is given by the partition function on a cylinder $\cal C$ slit along $(\ell,\tau)$, divided by the partition function on the full cylinder. 
Now consider the overlap
\begin{equation}\label{ratio}
\frac{{\rm Tr}\,(\rho_\ell\cdot\tilde\rho_\beta)}{\big(({\rm Tr}\,\rho_\ell^2)({\rm Tr}\,\tilde\rho_\beta^2)\big)^{1/2}}
=\frac{Z({\cal S}\oplus{\cal C})}{\big(Z({\cal S}\oplus{\cal S})Z({\cal C}\oplus{\cal C})\big)^{1/2}}\,,
\end{equation}
where, for example, $Z({\cal S}\oplus{\cal C})$ is the partition function on $\cal S$ sewn onto $\cal C$, such that the bottom edge of the slit in $\cal S$ is sewn to the top edge of $\cal C$ and vice versa (see Fig.~1). (This construction is similar to that used in formulating the R\'enyi-2 entropy as a path integral.) Note that in the ratio in (\ref{ratio}) the normalization factors all cancel.
\begin{figure}\label{SC}
\centering
\includegraphics[width=0.8\textwidth]{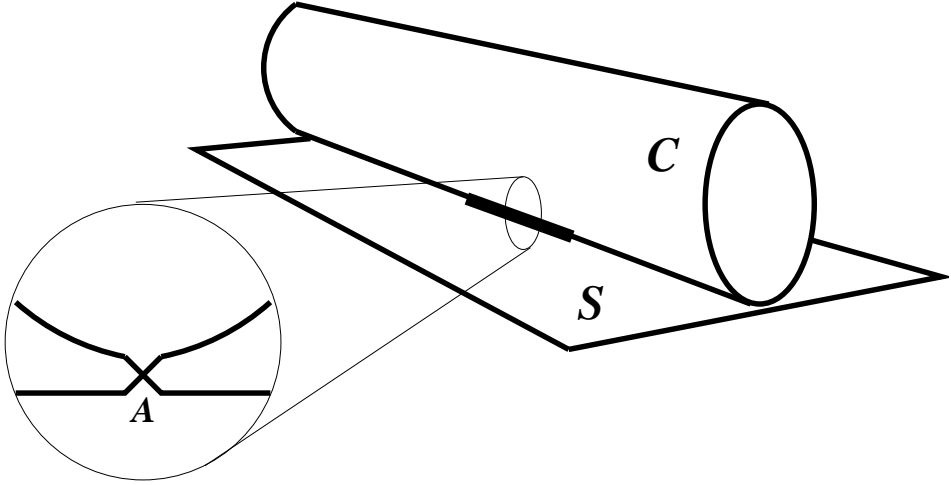}
\caption{The surface $S\oplus C$ in the numerator of (\ref{ratio}). The cylinder $C$ and the strip $S$ are sewn together along $A$ as shown in the inset.}
\end{figure}

The surfaces in the three factors in (\ref{ratio}) are all conifolds: the metric is euclidean except at the ends of the interval $\ell$ where they have a conical points with deficit angle $4\pi-2\pi$. In a CFT we may view their partition functions as correlators of \em twist operators \em $\cal T$ evaluated in the product of the CFTs on each component \cite{twist}. Thus
\begin{equation}\label{ratio2}
\fl\frac{Z({\cal S}\oplus{\cal C})}{\big(Z({\cal S}\oplus{\cal S})Z({\cal C}\oplus{\cal C})\big)^{1/2}}
=\frac{\langle {\cal T}(x_1,\tau){\cal T}(x_2,\tau)\rangle_{{\cal S}\otimes {\cal C}}}
{\big(\langle{\cal T}(x_1,\tau){\cal T}(x_2,\tau)\rangle_{{\cal S}\otimes {\cal S}}\langle{\cal T}(x_1,\tau){\cal T}(x_2,\tau)\rangle_{{\cal C}\otimes {\cal C}}\big)^{1/2}}\,,
\end{equation}
where now, for example, $\langle {\cal T}(x_1,\tau){\cal T}(x_2,\tau)\rangle_{{\cal S}\otimes {\cal C}}$ means the twist correlator on 
a direct product of the CFT on $\cal S$ with that on $\cal C$.

These twist operators behave in many respects like other semi-local operators in a CFT. In particular their correlators enjoy similar analyticity properties. Therefore we may take over the earlier arguments applied to correlators of ordinary operators. Applying the conformal mapping (\ref{zw}), they are related to correlators of twist operators on CFTs on 
${\mathbb H}\otimes{\mathbb C}, {\mathbb H}\otimes{\mathbb H},{\mathbb C}\otimes{\mathbb C}$ respectively.
Once the points $(x_1,\tau)$ and $(x_2,\tau)$ have fallen into a horizon, the cross-ratio $\eta_{12}$ is $\ll1$, and we may apply the OPE.

The OPE of twist operators has been discussed in \cite{CCT,Head}. In general it may be expressed as a sum of products of local operators in each sheet. In the present case
$$
{\cal T}(z_1)\cdot{\cal T}(z_2)=\sum_{k_1,k_2}C_{k_1,k_2}\Phi_{k_1}\Phi_{k_2}\,,
$$
where $\Phi_{k_1}$ and $\Phi_{k_2}$ are a complete set of operators, arranged in order of increasing dimension, in copies of the CFT on $(\mathbb H,\mathbb C),(\mathbb H,\mathbb H),(\mathbb C,\mathbb C)$ respectively, and the $C_{k_1,k_2}$ are (calculable) coefficients. The leading term comes from taking $\Phi_{k_1}=\Phi_{k_2}={\bf 1}$, so that, once the ends of the interval have fallen inside the horizon, the right hand side of (\ref{ratio2}) is asymptotically unity. The leading correction comes from the lowest dimension non-trivial operators $(\Phi_{k_1},\Phi_{k_2})$ which have non-zero expectation values. Since these are expected to be primary, and such operators have vanishing expectation value in $\mathbb C$, the leading corrections are only in the first factor in the denominator of (\ref{ratio2}), corresponding to the most relevant operators which have a non-vanishing expectation values on $(\mathbb H,\mathbb H)$ with the particular boundary state $|B\rangle$, as was found above. This gives the main result (\ref{close}) of this section. 

Note that if we use
the same method to compute the overlap between two thermal ensembles with different inverse temperatures $\beta$ and $\beta'$, the leading contribution to the corresponding ratio of partition functions (coming from the propagation of the ground state along the length $\ell$ for which the two cylinders are joined to give a single cylinder of circumference
$\beta+\beta'$) is
$$
\frac{\exp(\pi c\ell/6(\beta+\beta'))}{\left[\exp(\pi c\ell/6(2\beta'))\cdot\exp(\pi c\ell/6(2\beta))\right]^{1/2}}=
\exp\left(-\frac{\pi c\ell(\beta'-\beta)^2}{24\beta\beta'(\beta+\beta')}\right)
$$  
The overlap is therefore exponentially suppressed if $\beta'\not=\beta$ and $\ell\gg\beta,\beta'$. This shows, by the triangle inequality, that $\rho_\ell$ cannot be close to a thermal ensemble with inverse temperature $\beta'\not=\beta$.

\section{Quench from a more general boundary state and the GGE}\label{sec3}
In the section we consider a quench in a pure CFT from a more general state given by (\ref{generalstate}), which we rewrite as
\begin{equation}\label{eqphib}
|\psi_0\rangle\propto e^{-(\beta/4)H_{CFT}}\,\prod_ke^{-\beta_k\int\widetilde\Phi_k(x)dx}|B\rangle\,,
\end{equation}
where the product is now over all boundary operators excepting the stress tensor. Note that we have commuted $e^{-(\beta/4)H_{CFT}}$ through all the other terms so it stands on the left. Since commutators of $H$ with a given local operator generate others, the effect of this is to modify the couplings $\beta_k$.  
Although this expression is motivated by the idea that we should include only irrelevant boundary operators with scaling dimension $\Delta_k>1$, as we argue below it makes sense in some cases also to include relevant ones. 

As it stands, the expression (\ref{eqphib}) is only formal. If expanded in powers of the $\beta_k$ it will involve integrals of boundary correlators of the $\widetilde\Phi_k$ which will give rise to UV divergences if $\Delta_k>1$. These may be dealt with in the usual way by first imposing a UV cut-off in space, identifying the UV divergent terms by using the OPE, and then adding counterterms coupling to the higher dimension operators in this OPE in such a way as to cancel the divergences as the cut-off is removed. This has the effect that the renormalized $\beta_k$ appearing in (\ref{eqphib}) then depend on the bare values in a complicated non-linear fashion. 

This perturbative expansion should make sense for all irrelevant boundary operators for which $\Delta_k>1$. For relevant operators one should expect to encounter infrared divergences signaling the crossover to a new boundary fixed point. However in this case the finite effective inverse temperature provides a cut-off, and the expansion should still make sense as long as $\beta_k^{-1/(1-\Delta_k)}\gg\beta$, that is $\beta_k\ll\beta^{-(1-\Delta_k)}$.

Let us first consider the case when the $\widetilde\Phi_k$ are all descendants of the identity operator, plus their hermitian conjugates. These can be written as linear combinations of all possible powers of the components of the stress tensor 
$(T,\overline T)$ and its derivatives. In fact, since the conformally invariant boundary state satisfies
$$
\big(T(x)-\overline T(x)\big)|B\rangle=0\,,
$$
there is no need to distinguish between $T$ and $\overline T$ on the boundary. We denote these powers and derivatives of $T$ by $T^{(k)}(x)$. They should be normal-ordered as usual by point-splitting and removing the singular terms in the OPE.
The $T^{(k)}$ have the property that their boundary scaling dimensions $\Delta_k$ are integers $\geq2$. In fact, they should be even integers for the state in (\ref{eqphib}) to be invariant under reflection $x\to -x$. Note that there is in general more than one such descendant operator with dimension $\Delta_k$. 

As remarked above, there is no need to distinguish between $T^{(k)}(x)$ and $\overline T^{(k)}(x)$ on the boundary. However they may each be viewed as the boundary values of holomorphic and antiholomorphic operators $T^{(k)}(w)$ and $\overline T^{(k)}(\bar w)$ which are different in the bulk. In the path integral in the euclidean slab geometry, each exponential factor in (\ref{eqphib}) corresponds to inserting a term 
$$
\beta_k\int T^{(k)}(x)dx=\ffrac12\beta_k\int T^{(k)}(w)dw+\ffrac12\beta_k\int\overline T^{(k)}(\bar w)d\bar w
$$
along contours just above and below the lower and upper boundaries $\tau=\pm\beta/4$. There is a contour for each $k$,
and they should be ordered in increasing $k$. The main point now is that, because of the (anti-)holomorphicity, each of these contours may be freely deformed into the bulk, as long as they do not cross each other or the arguments of local observables in correlation functions. This corresponds to the statement that the charges which are the space integral of the $T^{(k)}$ commute with the hamiltonian. On transforming to the $z$-plane, the insertion becomes (for each $k$)
$$
\ffrac12\beta_k(-i\beta/\pi))^{1-\Delta_k}\int T^{(k)}(z)z^{\Delta_k-1}dz+ \cdots +{\rm c.c.}\,,
$$
summed over two contours: one from $z=0$ to positive real $+\infty$, the other from $z=0$ to real $-\infty$. The omitted terms in the above correspond to the contributions of other (holomorphic) descendants, which appear because the $T^{(k)}$ are not primary. However these will disappear when we reverse the mapping going to the cylinder.

We now continue the arguments of the local observables to real time. As before, we argue that at late times the $\bar z_j$ move off to $-i\infty$ so that effectively the boundary disappears. However, the boundary perturbations given by the above contour integrals remain. In the absence of the boundary we can now rotate one contour into the other, for example choosing them both to run from $z=0$ to $-i\infty$. This gives rise to a relative factor
\begin{equation}\label{factor}
(e^{-i\pi/2})^{\Delta_k}+(e^{i\pi/2})^{\Delta_k}=2\cos(\pi\Delta_k/2)\,.
\end{equation}
These manipulations are illustrated in Fig.~2.
\begin{figure}\label{contours}
\centering
\includegraphics[width=0.8\textwidth]{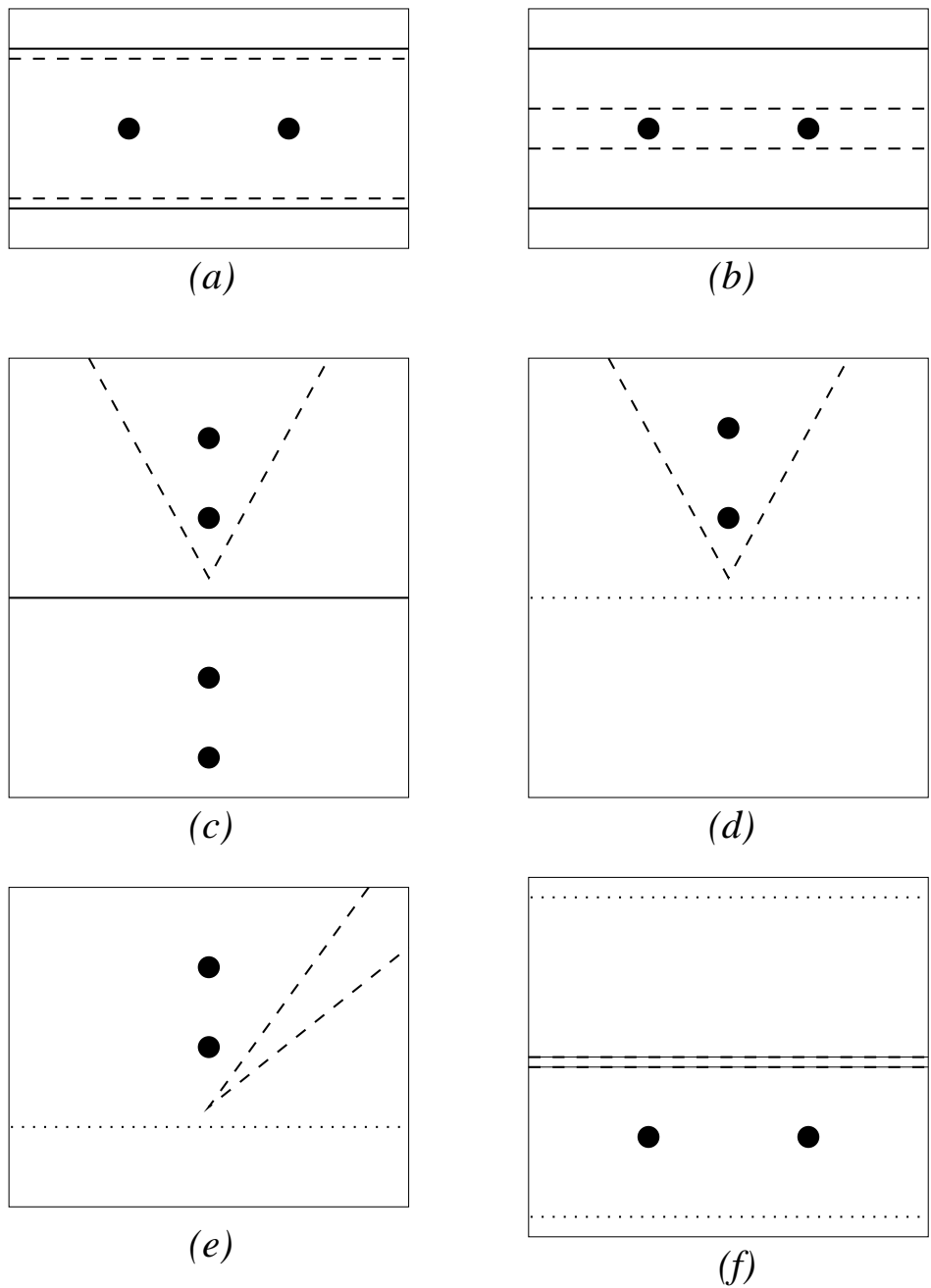}
\caption{The contour manipulations leading to (\ref{SGGE}). \em(a) \em initially  the perturbing operators are integrated along the dashed lines adjacent to the upper and lower boundaries of the strip; \em(b) \em  using (anti-)holomorphicity they are moved into the bulk;
\em(c) \em the strip is mapped to the upper half-plane: the correlator of local operators (indicated by black circles) depends on those points and their images; \em(d) \em after continuation to real time and evolving until both operators fall within the horizon, the image points move off to infinity: at this point the boundary (now indicated by a dotted line) effectively disappears and the correlator is the same as in the full plane; \em(e) \em the two parts of the dashed contour may now be rotated together and their contributions (which differ by a phase) combined; \em(f) \em on reversing the conformal mapping, the full plane becomes a cylinder with the insertion of a defect line along which the conserved currents are integrated.  }
\end{figure}

Although $\Delta_k$ is, so far, an integer, we have kept the more general expression for later use. Note that only even integer dimensional charges contribute: the odd ones are odd under parity and are zero acting on the initial state.

On transforming back to the $w$-plane (which is now a cylinder) we find an insertion in the path integral action
$$
\sum_k\beta_k\cos(\pi\Delta_k/2)H^{(k)}\,,
$$
where
$$
H^{(k)}=\int \big(T^{(k)}(x,\tau)+ \overline T^{(k)}(x,\tau)\big)dx\,,
$$
along $\tau=\beta/2$. However, because $H^{(k)}$ commutes with the hamiltonian, this contour could be along any constant imaginary time line which does not separate any of the arguments $w_j$ of the observables. In addition, the contours for different $k$ should be correctly ordered, reflecting the fact that the $H^{(k)}$ do not in general commute among themselves. 

Thus the correlation functions at times after all the points have fallen within the same horizon are given by a path integral with weight
\begin{equation}\label{SGGE}
e^{-S_{CFT}-\sum_k\beta_k\cos(\pi\Delta_k/2)H^{(k)}}\,,
\end{equation}
where the first term is the usual action for the CFT. This is just the path integral formulation of the (non-Abelian) GGE. 
Note that odd-dimensional charges do not contribute. 

\subsection{Physical consequences of the GGE.}
In what ways does the post-quench GGE differ from the Gibbs ensemble in measurable quantities, for example the correlation functions of local observables? Recall that in a Gibbs ensemble in a CFT, the stationary 2-point function of any scaling operator decays as
$$
\langle\Phi(x_1,t)\Phi(x_2,t)\rangle_\beta\sim e^{-2\pi\Delta_\Phi|x_1-x_2|/\beta}\,,
$$
for $|x_1-x_2|\gg\beta$, where  $\Delta_\Phi$ is the overall scaling dimensions of $\Phi$ (assumed to be a scalar). The reason for this is that the
generator of translations along the cylinder is
$$
h=(1/2\pi)\int T_{xx}d\tau=(1/2\pi)\int (T+\overline T)d\tau=(2\pi/\beta)\big(l_0+\bar l_0-(c/12)\big)\,,
$$
where 
\begin{equation}\label{Fourier}
\fl T(\tau)=(2\pi/\beta)^2\sum_q(l_q-\ffrac c{24}\delta_{q0})e^{2\pi iq\tau/\beta}\,,\quad \overline T(\tau)=(2\pi/\beta)^2\sum_q
(\bar l_q-\ffrac c{24}\delta_{q0})e^{-2\pi iq\tau/\beta}
\end{equation}
(we use lower-case letters to distinguish these from the hamiltonian $H$ and the associated Virasoro generators $L_q,\overline L_q$.) The eigenvalues of $l_0+\bar l_0$ are the total scaling dimensions $\Delta_\Phi$. 

In the GGE, the additional terms contribute to the action 
$$
\sum_k\beta_k\cos(\pi\Delta_k/2) \int_0^L \big(T^{(k)}(x,\tau_k)+\overline T^{(k)}(x,\tau_k)\big) dx\,. 
$$
where the $\tau_k$ are ordered in increasing $k$.

In terms of the generator of translations along the cylinder, there is now an additional term
\begin{equation}\label{deltah}
h\to h+\delta h= h+\beta^{-1}\sum_k\beta_k\cos(\pi\Delta_k/2) \int_0^\beta\big(\!\no{T^{(k)}(\tau_k+\tau)}+\no{\overline T^{(k)}(\tau_k+\tau)}\!\!\big)d\tau\,.
\end{equation}
We have averaged over the overall center of mass of the $\{\tau_k\}$ to emphasize the stationarity of the ensemble. This expression may be written in terms of the $l_q,\bar l_q$.

As an example, the lowest (dimension 4) operators correspond to $\no{T^2}$ and $\no{\overline T^2}$.
Inserting the Fourier decomposition (\ref{Fourier}), we have, before normal ordering
$$
\delta h= \beta_2(2\pi/\beta)^4\sum_q(l_q-\ffrac c{24}\delta_{q0})(l_{-q}-\ffrac c{24}\delta_{-q0})+{\rm c.c.}
$$
In this case, the effect of normal ordering is to move all the $l_q$ with $q>0$ to the right. Subtracting off the singular terms in the OPE corresponds to neglecting the commutator when we do this.
Thus
$$
\delta h=\beta_2(2\pi/\beta)^4\left (2\sum_{q>0}l_{-q}l_q+(l_0-\ffrac c{24})^2\right)+{\rm c.c.}
$$
The terms with $q>0$ annihilate any highest weight state, corresponding to the ground state of $h$ or the insertion of any primary operator at infinity.
Thus the partition function of the GGE is now
$$
Z\sim\exp L\left((\pi c/6\beta)-2\beta_2(2\pi/\beta)^4(c/24)^2\right)\,.
$$
Thus the energy density $-(1/L)\partial_\beta\log Z$ is modified, and it is no longer simply proportional to the entropy density $(1/L)\big(\beta\partial_\beta-1\big)\log Z$. 

The 2-point function of a scalar primary operator now decays with a correlation length
\begin{equation}\label{xiGGE}
(\xi^\Phi_{GGE})^{-1}=(2\pi\Delta_\Phi/\beta)+\beta_2(2\pi/\beta)^4\big((\Delta_\Phi-\ffrac c{12})^2-(\ffrac c{12})^2\big)\,.
\end{equation}
Thus in the GGE, the effective temperature as measured thermodynamically, or by a correlation function, and therefore in terms of fluctuations, depends on the observable.

Higher dimension operators may be dealt with in a similar way, although the algebra becomes more complicated. The form of the next few normally ordered operators has been given in Ref.~\cite{Zam}. The general picture does not change: the correlation length in the GGE will be expressed as a power series in the scaling dimensions of the particular observable.

Note that this will also have an effect on dynamical quantities. Since the time evolution remains governed by $H_{CFT}$, there will still be a sharp horizon (rounded on the scale of $\beta$), and the spatial decay of a 2-point function in the stationary state for $|x_1-x_2| <2vt$ should match across the horizon with the time decay of the product of the 1-point functions for
$|x_1-x_2|>2vt$. Thus a 1-point function (which does not vanish in the initial state) should decay as 
\begin{equation}\label{1pt}
\langle\Phi(x,t)\rangle\sim e^{-vt/\xi^\Phi_{GGE}}\,.
\end{equation}
This has been verified for the Ising model \cite{Ising} with an explicit form for $\xi_{GGE}$.

\subsection{Perturbative derivation.}
The above argument relies on distorting space-time contours in the exponential weighting of the path integral. Since this is somewhat heuristic, it is therefore reassuring to see that similar consequences can be seen within straightforward perturbation theory in the parameters $\beta_k$. 

We illustrate this by considering the modified decay of the above 1-point function (\ref{1pt}), and the simplest example of a $T^2$ operator perturbing the initial state. This may be evaluated by continuing to imaginary $\tau$ the sum of integrated correlators 
$$
\sum_{n=0}^\infty(1/n!)\int du_1\ldots du_n\langle\Phi(0,\tau)\,T^2(u_1,\pm\ffrac\beta4)\ldots T^2(u_n,\pm\ffrac\beta4)\rangle du_1\ldots du_n\,,
$$
where the integrals are along the upper and lower edges of the strip. (As discussed above, these must be regulated with a UV cut-off and the divergent terms subtracted.) The correlations functions are in the CFT with the unperturbed boundary state. As such, they may be estimated by transforming to the $z$-plane $\mathbb H$ as before. The correlator depends on $(0,\tau)$ and its image, which as before maps onto the imaginary $z$-axis, although not at conjugate points once the continuation to real time is made. 
The points $u_j$ map onto points $x_j=\pm e^{(2\pi/\beta)u_j}$ on the real axis. 

Once again, points separated by distances $>O(\beta)$ in the $w$-plane will become exponentially ordered in the $z$-plane, and, apart from these cross-over regions, we may apply the OPE.   

Let us consider the case $n=1$ as an example. This involves the 2-point function of the bulk operator $\Phi$ with the boundary operator $T^2$. Since the form of this follows from global conformal invariance, at this order we may initially consider a more general boundary operator $\widetilde\Phi_2$, with scaling dimension $\Delta_2$, rather than $T^2$.

The first order correction is 
$$
-\beta_2\int\langle\Phi(0,\tau))\widetilde\Phi_2(u,\pm\ffrac\beta4)\rangle du\,.
$$
As before we conformally map this into the upper half plane by $z=ie^{2\pi w/\beta}$, giving
$$
-\beta_2 b_{\Phi\widetilde\Phi_2}\int_0^\infty|dw/dz|^{-2\Delta_\Phi}|du/dx|^{1-\Delta_2}\frac{dx}{(i(\bar z- z))^{2\Delta_\Phi-\Delta_2}(z-x)^{\Delta_2}(\bar z-x)^{\Delta_2}}\,,
$$
plus a similar contribution from the other boundary, which means replacing $x\to-x$. We have used the explicit form of the 2-point function in $\mathbb H$ (which has the same form as the 3-point function in full plane.) Here $b_{\Phi\widetilde\Phi_2}$ is a bulk-boundary OPE coefficient. Continuing $\tau\to it$ as before, we have $|dw/dz|\sim(\beta/2\pi), |du/dx|\sim(\beta/2\pi x)$, giving

$$
-\beta_2 b_{\Phi\widetilde\Phi_2}(2\pi/\beta)^{2\Delta_\Phi-\Delta_2}(e^{2\pi t/\beta})^{\Delta_2-2\Delta_\Phi}
\int_{-\infty}^\infty\left(\frac{e^{2\pi u/\beta}}{(e^{2\pi u/\beta}-ie^{-2\pi t/\beta})(e^{2\pi u/\beta}+ie^{2\pi t/\beta})}\right)^{\Delta_2} du\,,
$$
plus the complex conjugate contribution from the other boundary. 

For $u<-t-O(\beta)$ we may approximate the integral giving
$$
-\beta_2 b_{\Phi\widetilde\Phi_2}(2\pi/\beta)^{2\Delta_\Phi-\Delta_2}(e^{2\pi t/\beta})^{\Delta_2-2\Delta_\Phi}\int_{-\infty}^{-t}(e^{2\pi u/\beta})^{\Delta_2} du=
-\beta_2 b_{\Phi\widetilde\Phi_2}(2\pi/\beta)^{\Delta_\Phi-\Delta_2}(e^{2\pi t/\beta})^{-2\Delta_\Phi}\,.
$$
We get the same contribution from $u>t$, and from the other boundary for $|u|>t$. These all give an $O(\beta_2)$ correction to the pre-factor of the 1-point function.

However, from $-t<u<t$ we have
$$
-\beta_2 b_{\Phi\widetilde\Phi_2}(2\pi/\beta)^{2\Delta_\Phi-\Delta_2}(e^{2\pi t/\beta})^{\Delta_2-2\Delta_\Phi}
\int_{-t}^t(ie^{2\pi t/2\beta})^{-\Delta_2}du\,.
$$
Adding the contribution from the other boundary we get 
$$
-\beta_2 b_{\Phi\widetilde\Phi_2}\cos(\pi\Delta_2/2)\,t\,(2\pi/\beta)^{2\Delta_\Phi-\Delta_2}(e^{2\pi t/\beta})^{-2\Delta_\Phi}\,,
$$
so, relative to the $O(\beta_2^0)$ term, we get a factor
\begin{equation}\label{linear}
1-\beta_2 b_{\Phi\widetilde\Phi_2}\cos(\pi\Delta_2/2)(2\pi/\beta)^{-\Delta_2}\,t\,.
\end{equation}

Therefore the first-order correction appears to give a larger contribution as $t\to\infty$ than the zeroth-order term. The reason that it is $\propto t$ is physically clear from the quasiparticle picture: at time $t$ the perturbation of the initial state is felt only from regions with $|u|<t$, by causality. Within this region, however, the integrand is approximately constant. Therefore the contribution is $\propto t$. For the same reason, we then expect the leading behavior of higher-order contributions to behave like $(-\beta_2)^nt^n/n!$, so that the first-order correction simply exponentiates. This can be seen more clearly in the $z$-plane: the leading contribution now comes from the region where $|z|\ll x_1\ll x_2\ll\cdots\ll x_n\ll|\bar z|$. In this limit we may use the bulk-boundary OPE repeatedly, each time getting a factor $b_{\Phi\widetilde\Phi_2}$. Note that the factor $\cos(\pi\Delta_2/2)$ in (\ref{GGE}) also naturally arises in this approach. 

Specializing to the case $\widetilde\Phi_2=T^2$, we have by the conformal Ward identity that $b_{\Phi,T^2}=\Delta_\Phi^2$, which then confirms the term $O(\Delta_\Phi^2)$ in (\ref{xiGGE}). (The other term $\propto c$ arises from normal-ordering.) 

The result (\ref{linear}) and its exponentiation have a simple interpretation in the quasiparticle picture:
the perturbing operator acting on the initial state gives rise to additional quasiparticle pairs. Only those coming from the past light cone of $(x,t)$ affect $\langle\Phi(x,t)\rangle$, and those coming from initial points $(x,0)$ separated by distances $>\beta$ are incoherent, so their effects exponentiate.

\subsection{Non-integer exponents and semi-local conservation laws}
The above argument assumed that the operators $\widetilde\Phi_k(x)$ acting at the boundary have (even) integer dimension.
However, in most CFTs and for most conformal boundary states there will also exist operators with non-integer dimensions. One may ask whether these also lead to a GGE in the stationary state. For the above argument to apply, we have to be able to think of these boundary operators in some sense as the limits to the boundary of bulk holomorphic and anti-holomorphic operators $\Phi_k(z)$ and $\overline\Phi_k(\bar z)$. We argue that, at least for rational CFTs, this is indeed the case. Recall that in this case all primary operators (under Virasoro or some extended algebra) have descendant null states which implies that their boundary correlators $\langle\widetilde\Phi_k(x_1)\widetilde\Phi_k(x_2)\ldots\rangle$ satisfy systems of linear differential equations with singular coefficients at coincident points. The solutions then have power law singularities at these points, but may otherwise be continued uniquely into the upper (or lower) half-plane as holomorphic (or anti-holomorphic) functions. These may be viewed as the (multiple-valued) correlators of holomorphic (or anti-holomorphic) operators. If the boundary operator has boundary scaling dimension $\Delta_k$, these bulk operators have dimensions $(\Delta_k,0)$ and $(0,\Delta_k)$ respectively. 

Another way to construct these bulk operators, which works in principle for any CFT, is to cut out a small disc of radius $\epsilon$ around a bulk point $(z,\bar z)$, then insert $\widetilde\Phi_k$ at the point $z+\epsilon e^{i\theta}$ on the boundary of the disc. Correlators of the bulk holomorphic operator are then defined as weighted integrals of those of the boundary operator:
$$
\langle\Phi_k(z)\ldots\rangle=\lim_{\epsilon\to0}\int_0^{2\pi}e^{-i\Delta_k\theta}\langle\widetilde\Phi_k(z+\epsilon e^{i\theta})\ldots\rangle d\theta\,.
$$
This resembles the construction of discretely holomorphic bulk operators in lattice loop models \cite{latticeholo}.

Having defined these bulk operators, the previous construction of the GGE follows, at least formally. However the associated charges $Q_k=\int \Phi_k(u,\tau)du+\int\overline\Phi_k(u,\tau)du$ do not quite commute with the hamiltonian $H_{CFT}$. This is because if we consider a closed contour which is the boundary of a long rectangle $(-\frac12L<u<\frac12L,\tau_1<\tau<\tau_2)$, the contributions from the end pieces do not cancel for periodic boundary conditions since they carry different phases $e^{\pm i\Delta_k}$. They would in fact cancel for suitable twisted boundary conditions on the fields of the theory. Thus we may think of the action of a single $Q_k$ as switching the boundary conditions between different sectors of the theory. It is a parafermionic charge. If $\Delta_k$ is rational, with lowest denominator $M$, it is only after acting $M$ times that we return to the original boundary conditions. Thus, in the GGE expression
$$
{\rm Tr}\, e^{-\beta H_{CFT}}\prod_ke^{-\beta_k\cos(\pi\Delta_k/2)Q_k}\,,
$$
the trace is supposed to project onto only those terms in the expansion of the exponential containing integer powers of $\beta_k^M$. 

\subsubsection{Example: the transverse Ising model in a longitudinal field.}
As an example, we consider the CFT description of a quench in the Ising chain with hamiltonian
$$
H_{\rm Ising}=-J\sum_j\sigma^z_j\sigma_j^z-h_x\sum_j\sigma_j^x-h_z\sum_j\sigma_j^z\,,
$$
from the ground state with $h_x\gg J$ and small longitudinal field $h_z\ll J$ to the critical point $h_x=J$, $h_z=0$.
This has not been considered in the literature because the model is no longer solvable in terms of free fermions when $h_z\not=0$.
The appropriate conformal boundary state when $h_x\gg J$ and $h_z=0$ is that corresponding to free boundary conditions on the $\sigma_j^z$. This state supports one primary operator of scaling dimension $\delta=\frac12$ which is interpreted as the scaling limit of the local magnetization $\sigma^z$. Turning on a small $h_z$ is equivalent to perturbing the boundary state as described above. Although in this case the perturbation is relevant, as was argued above, we may still consider it as small as long as $h_z\ll \beta^{-1/2}\sim(h_x/J-1)^{1/2}$. Note that the expansion in terms of correlators of $\sigma^z$ on the boundary contains only even powers of $h_z$ because of the ${\mathbb Z}_2$ symmetry of the state. 

The holomorphic and anti-holomorphic extensions of this boundary operator are in this case well understood -- they are nothing but the fermions $(\psi(z),\bar\psi(\bar z))$ of the bulk Ising CFT. Thus the GGE in the case should contain fermonic charges $(Q,\overline Q)$ which, as is well known, act to switch between periodic and anti-periodic boundary conditions on the Ising spins $\sigma^z$. If we consider a fixed choice of boundary conditions, for example periodic, only even powers of these charges enter the computation of correlation functions of local spins in this GGE. This is completely consistent with the fact that only even powers of $h_z$ enter the perturbative expansion.

\section{Perturbations of the CFT}\label{sec4}
In this section we consider what happens in the more realistic case when the time evolution is governed by a hamiltonian $H$ which differs from that of a pure CFT by the addition of couplings to irrelevant operators. While these give only corrections to scaling at the zero-temperature critical point, their precise influence on the quench dynamics has not so far been investigated. Studies in integrable models suggest that the quasi-particle picture remains valid, with due allowance for the different dispersion relation.

While it is possible to study these effects perturbatively in the couplings to these irrelevant operators, similarly to the case of the deformed initial state in Sec.~\ref{sec3}, this similarly breaks down at late times, and, unlike the former case, it is not obvious to see how to resum the series. Instead we take a different approach, which, as we will show, agrees with perturbation theory at low orders. Unfortunately this approach only works for irrelevant operators which are descendants of the identity, that is, polynomials in the components $(T,\overline T)$ of the stress tensor and their derivatives. In fact, most of the analysis will be directed at the dimension 4 operators $T^2$, $ \overline T^2$ and $T\overline T$, arguing later that the general picture should remain valid for higher dimensional operators if weakly coupled. However the dimension 4 operators are important, as they are generated in almost any quantum critical hamiltonian with a linear low-energy dispersion relation by integrating out higher energy degrees of freedom up to the scale of the UV cut-off.\footnote{In a free theory, with has conserved $U(1)$ currents $(J,\overline J)$, the curvature of the dispersion relation also results in terms like $\bar\partial\overline J\partial J$.} The first two terms break Lorentz invariance, and would be generated, for example, in a lattice model. However, as discussed in the previous section, their space integrals in fact commute with the hamiltonian and they preserve the integrability of the model. However the third term, while preserving relativistic invariance, introduces right-left scattering and does not commute with $H_{CFT}$.\footnote{However it may still be integrable, see for example \cite{Zam2}.} This may be seen by quantizing the CFT on a circle of length $L$, whence
$$
\int_0^L T^2(x)dx\propto \sum_kL_kL_{-k},\quad\int_0^L\overline T^2(x)dx\propto \sum_k\overline L_k\overline L_{-k},
\quad \int_0^L T(x)\overline T(x)dx\propto \sum_kL_k\overline L_k\,.
$$
The first two commute with $H_{CFT}\propto L_0+\overline L_0$, while the last does not. 

Let us therefore consider the perturbed hamiltonian
$$
H=H_{CFT}-\ffrac12\lambda\int(T^2+\overline T^2)dx-\mu\int T\overline Tdx\,,
$$
which corresponds to a euclidean action
$$
S^E=S^E_{CFT}-\ffrac12\lambda\int(T^2+\overline T^2)dxd\tau-\mu\int T\overline Tdxd\tau\,.
$$
Note that we do not normal order, but will shortly introduce an explicit UV regulator. The coupling constants $(\lambda,\mu)$ have the dimensions $({\rm length})^2$.
The additional terms, quadratic in $T$ and $\overline T$, may be generated by the usual trick of introducing auxiliary complex conjugate fields $(\bar\xi,\xi)$, with spins $\pm2$, in the functional integral:
\begin{equation}\label{SE}
\fl S^E=S^E_{CFT}+\frac1{\mu^2-\lambda^2}\int\big(\mu\bar\xi\xi-\ffrac12\lambda(\bar\xi^2+\xi^2)\big)dxd\tau+\int(\bar\xi T+\xi\overline T)dxd\tau
\end{equation}
In Feynman diagram terms the interaction may be thought of as proceeding by the exchange of $(\bar\xi,\xi)$ `particles' which couple linearly to $T$ and $\overline T$. A UV regulator may be introduced by adding a kinetic term 
$\propto\int(\partial\xi)(\bar\partial\bar\xi)dxd\tau$. However, this makes sense as a unitary theory in Minkowski space only if the eigenvalues  of the (mass)$^2$ matrix $(\mu\pm\lambda)^{-1}$ are positive. One might assert the freedom to choose the signs differently, but these would not result from the integration over unitary relativistically invariant higher energy fluctuations, at least of a scalar nature. As we shall see, the choice of the `wrong' signs would lead to superluminal speeds of propagation in the low-energy effective theory. This, of course, is not a problem if this theory is describing a lattice model which does not have to possess a relativistic ultraviolet completion.



Although we shall continue to work in euclidean space for the time being, continuing to real time only at the end, it is interesting to note that in Minkowski space (\ref{SE}) corresponds to 
$$
S^M=S^M_{CFT}+\ffrac12\lambda\int(T_{++}^2+T_{--}^2)dxdt+\mu\int T_{++}T_{--}dxdt\,,
$$
where we have introduced light-cone coordinates $x^{\pm}=x\pm t$. It is instructive to consider the case of a free scalar field $\phi$, for which $T_{\pm\pm}\propto(\partial_\pm\phi)^2$. In momentum space the interaction vertices are
then $-i\lambda(k_{1+}k_{2+}k_{3+}k_{4+}), -i\lambda(k_{1-}k_{2-}k_{3-}k_{4-}),-i\mu(k_{1+}k_{2+}k_{3-}k_{4-})$.
$\mu\pm\lambda>0$ then corresponds to an \em attractive \em interaction. (This may also be seen by comparing (\ref{SE}) with case of a simple non-derivative interaction $S=S_{free}-\mu\int\phi^4dxd\tau$. The `usual' sign with $\mu<0$ is repulsive.) 

Note that the couplings $\lambda,\mu$ have dimension (length)$^2$. In what follows we assume that they are small compared the dominant length scale of the problem which is $\beta^2$. This means that we can assume that the (dimensionless) fields 
$(\bar\xi,\xi)$ are $\ll1$. 

The main point now is that the last term in (\ref{SE}) may be reinterpreted, using the definition of the stress tensor,  as the response of the action to a small change in the metric $\delta\bar g=\delta g^{ww}=\bar\xi$, $\delta g=\delta g^{\bar w\bar w}=\xi$,
which themselves may be viewed as gaussian random fields with covariance
\begin{eqnarray*}
{\mathbb E}[\delta g(u,\tau)\delta g(u',\tau')]&=&\lambda\delta(u-u',\tau-\tau')\,,\\
{\mathbb E}[\delta\bar g(u,\tau)\delta\bar g(u',\tau')]&=&\lambda\delta(u-u',\tau-\tau')\,,\\
{\mathbb E}[\delta g(u,\tau)\delta\bar g(u',\tau')]&=&\mu\delta(u-u',\tau-\tau')\,.
\end{eqnarray*}
If we introduce a UV regulator as above, the delta functions are smeared over a length scale $O(\lambda^{1/2},\mu^{1/2})$. 

The deformed metric is
$$
ds^2=dwd\bar w+\delta gdw^2+\delta\bar gd\bar w^2\,.
$$
To lowest order (only) this is equivalent to a coordinate change
$ds^2=d\tilde wd\bar{\tilde w}$ where
$$
d\tilde w=dw+\delta \bar gd\bar w,\quad d\bar{\tilde w}=d\bar w+\delta gdw\,.
$$
Again to lowest order, this implies that the correlation function of any product of local observables evaluated with the perturbed action (\ref{SE}) is given by the expectation value over the random fields of the same product in the pure CFT at shifted values of the arguments:
$$
\langle{\cal O}(w_j,\bar w_j)\rangle_{S_{CFT}+\delta S}={\mathbb E}[\langle{\cal O}(\tilde w_j,\bar{\tilde w}_j)\rangle_{S_{CFT}}]\,.
$$
where ${\mathbb E}[\cdot]$ means the average over the gaussian fields.

We now apply these results to the quench scenario. Recall  that a point $(x_j,t_j)$ in Minkowski space is mapped to
$$
w_j=x_j-t_j,\quad \bar w_j=w_j'=x_j+t_j+i\beta/2\,,
$$
on the cylinder.
Thus the effect of the coordinate shift in light-cone coordinates is
$$
d\tilde x_+=dx_++\delta\bar g\,dx_-\,,\quad d\tilde x_-=dx_-+\delta g\,dx_+\,.
$$
Note that we have avoided going into Minkowski signature until this last step: although a similar argument should be possibly directly in real time (thereby being applicable to the effect of the perturbation on more general time evolution besides that of quench scenario) in practice we have been unable to resolve tricky questions of reality in the gaussian measure.

We now discuss some simple consequences of this picture.

\subsection{Renormalization of propagation speed.}
In the shifted (tilded) coordinates the correlations are to be evaluated in the pure CFT, so signals propagate with speed 
$v$ (which so far we have taken to be 1.) The left- and right-moving null geodesics are given by $d\tilde x_+=0$, 
$d\tilde x_-=0$ respectively, that is in the original coordinates
\begin{equation}\label{dx}
dx_+=-\delta\bar g\,dx_-\,,\quad dx_-=-\delta g dx_+\,.
\end{equation}
The first effect is due to the fact that, in the quench scenario, $\delta g$ and $\delta\bar g$ have non-zero expectation values. This is because $\langle T\rangle$ and $\langle\overline T\rangle$ are both non-vanishing, see (\ref{H}).
From (\ref{SE}) we that see that at the saddle-point $\langle\delta g\rangle=\langle\xi\rangle=-(\mu+\lambda)\langle T\rangle\not=0$, and similarly for the complex conjugate. To this approximation (neglecting fluctuations), the
null geodesics become $dt=\pm(1+2\langle\delta g\rangle)dx$, so the average speed of propagation is renormalized.
Since $\langle T\rangle=\langle\overline T\rangle<0$ in the initial state (and thereafter), $\langle\delta g\rangle$ and $\langle\delta\bar g\rangle$ are both positive if $\mu+\lambda>0$, and so the speed is reduced.

More directly, we see from (\ref{SE}) that there are terms in the hamiltonian
$$
\delta H=-\int\big(\lambda\langle T\rangle T+\lambda\langle\overline T\rangle\overline T+\mu\langle\overline T\rangle T
+\mu\langle T\rangle\overline T\big)dx\,.
$$
Recalling that $\langle T\rangle=\langle\overline T\rangle=-\pi c/12\beta^2$, we see that this term is
$$
\delta H=(\lambda+\mu)(\pi c/6\beta^2)H\,.
$$
This is equivalent to a rescaling of the time coordinate $t\to t\big(1+(\lambda+\mu)(\pi c/6\beta^2)\big)$. On the other hand, a similar argument applied to the generator $h$ of translations along the cylinder or strip yields the opposite sign for the shift (because $T_{xx}=-T_{\tau\tau}$ in the CFT) which is equivalent to a rescaling $x\to x\big(1-(\lambda+\mu)(\pi c/6\beta^2)\big)$. The propagation velocity is therefore renormalized according to
$$
v\to v\big(1-2(\lambda+\mu)(\pi c/6\beta^2)\big)\,.
$$
If $\lambda+\mu>0$ this corresponds to a reduction, which is physically reasonable: the attractive interaction causes a positive time delay as the quasiparticles scatter from the rest of those in the finite density sea. Note that this attenuation is greater at higher effective temperatures as expected. If we had chosen the other signs for the couplings (corresponding to a non-unitary interaction with the higher energy scales) this would have led to an increase in propagation speed with increasing temperature. In a microscopically relativistic theory this would be impossible. Similar observations (of couplings with the wrong sign implying superluminal propagation) have been made in the context of unification in particle physics \cite{super}. 

\subsection{Diffusive broadening of the horizon}
We now return to (\ref{dx}) and include the short-range correlated random nature of $(\delta g,\delta\bar g)$. For convenience we assume that the rescaling of $x$ and $t$ implied by the fact that they have non-zero expectation values has already been performed. If we integrate (\ref{dx}) to lowest order in the couplings, the equations for a right-moving null geodesic starting at $(x_+=0,x_-=x_-(0))$ is
$$
x_-(x_+)=x_-(0)-\int_0^{x_+}\delta g(x_+',x_-(0))dx_+'\,.
$$
If the integrand were delta-correlated only in  $x_+'$, this would lead to Brownian motion for $x_-(x_+)$. However in fact
$$
\langle\delta g(x_+',x_-')\delta g(x_+'',x_-'')\rangle=\lambda\delta(x_+'-x_+'')\delta(x_-'-x_-'')\,,
$$
so that the effective diffusion constant will be divergent. Even if a UV regulator is used to modify the correlator of the random field, the result will depend strongly on this. Thus individual null geodesics will behave rather wildly on the scale of  the UV cut-off. 

However, recall that the horizons in the pure CFT have a width $O(\beta)$. If we define the mean $x_-$ coordinate of the horizon by the random variable
$$
X_-(x_+)=\beta^{-1}\int_{-\beta/2}^{\beta/2}x_-(x_+)dx_-(0)\,,
$$
this has variance
$$
\langle X_-(x_+)^2\rangle=(\lambda/\beta)x_+\,.
$$
(If we use a more realistic profile for the unperturbed horizon, the coefficient will change but not the functional dependence.) Thus we see that the effect of the $T^2$ and $\overline T^2$ operators is to diffusively broaden the horizon
to a width $O\big((\lambda t/\beta)^{1/2}\big)$. However this is still smaller than $t$ so on this scale the horizons remain well-defined. Note also that the $T\overline T$ term, which describes left-right scattering, does not contribute to this effect.
This can be seen in the example of the free boson discussed above. Since the scattering is proportional to 
$(k_{1+}k_{2+}k_{3-}k_{4-})$,
it vanishes on both left- and right-moving light-cones and so does not modify them. This is consistent with a pure $T\overline T$ interaction preserving exact Lorentz invariance.

\subsection{Correlation functions.}
We now discuss how the coupling to the random metric affects the asymptotics of correlation functions. First consider an equal-time correlator $\langle\Phi(x_1,t)\Phi(x_2,t)\rangle$ in the stationary state, that is when $2t-|x_1-x_2|\gg\beta$. In the unperturbed theory its decay is given by $(\xi^\Phi_\beta)^{-1}=2\pi\Delta_\Phi/\beta$. Consider the space-like geodesic between these two points. Along this we have
$$
d\tilde x=\ffrac12(d\tilde x_++d\tilde x_-)=dx+\delta\bar gdx_-+\delta gdx_+\approx dx+(\delta\bar g+\delta g)dx\,,
$$
since to this order $dt$ can be neglected. Thus
$$
\langle\Phi(x_1,t)\Phi(x_2,t)\rangle_{H+\delta H}\sim{\mathbb E}\left[e^{-2\pi\Delta_\Phi|\tilde x_1-\tilde x_2|/\beta}\right]\,,
$$
where the expectation on the right hand side is with respect to the gaussian random fields. From the above we see that $\tilde x_1-\tilde x_2$ is a Brownian motion, but, once again, because the fields $\delta g,\delta\bar g$ are defined at equal times, the effective diffusion constant depends strongly on the UV cut-off. Again the resolution is not to consider the $x$-cooordinate at a particular imaginary time $\tau$, but averaged over the cylinder. This is equivalent to computing the stationary behavior. This variable has a finite variance $2(\lambda+\mu)/\beta$, and so we find
$$
\langle\Phi(x_1,t)\Phi(x_2,t)\rangle_{H+\delta H}\sim 
\exp\left(-\left(\ffrac{2\pi\Delta_\Phi}\beta-\ffrac{(2\pi\Delta_\Phi)^2(\lambda+\mu)}{\beta^3}\right)|x_1-x_2|\right)\,.
$$
This result agrees with first-order perturbation theory, since $\langle\Delta_\Phi|T^2,\overline T^2,T\overline T|\Delta_\Phi\rangle=(2\pi/\beta)^4\Delta_\Phi^2$.

\subsection{Higher order effects.}
Finally we consider the effects of taking into account higher orders in $\lambda$ and $\mu$, as well as higher order descendants $T^{(k)}$ of the stress tensor in the quench hamiltonian. The latter terms can still be described by a coupling to a random metric with non-gaussian correlations, which should not change the overall picture if they are small. 

In general, since the metric is 2-dimensional, it is locally conformally flat, so it is always possible to choose a local system of coordinates $\tilde x_\pm$ so that the metric has the form
$$
ds^2=e^{\phi(\tilde x_+,\tilde x_-)}\,d\tilde x_+d\tilde x_-\,.
$$
However, in higher orders this will correspond to non-zero curvature $R\propto\partial^+\partial^-\phi$. This first arises at second order, when $R\propto\partial^+\partial^-(\delta g_+\delta g_-)$. Thus it first appears as a consequence of (potentially non-integrable) L-R scattering. In fact, pure LL and RR scattering is always equivalent to a simple coordinate change $x_\pm\to\tilde x_\pm$ and introduces no curvature. Thus in the latter case we expect to see only broadening of the horizon, possibly non-gaussian. On the other hand, strong negative curvature effects could lead to other effects such as the chaotic divergence of geodesics. However this seems difficult to analyze quantitatively. 

Note that while this discussion is reminiscent of 2d quantum gravity, the physics is very different -- the 'graviton' here has a heavy mass $O((\mu\pm\lambda)^{-1/2})$. In fact, the physics is more similar to the scattering of light by gravitationally bound dark matter in the distant universe. 

\section{Discussion.}

In this paper we have considered various extensions of the initial results in CC \cite{CC} concerning the quench to a critical point in a 1+1-dimensional CFT from a state with short-range correlations. First, we showed that once a finite interval falls inside the horizon its reduced density matrix becomes exponentially close in $L_2$ norm to that of a thermal state, generalizing the results of CC which applied only to correlators of primary operators. It would be straightforward to extend this to other $L_p$ norms using the same techniques. Second, we showed that deformations of the CC initial state by action of integrals of local boundary operators should lead to a (non-Abelian) GGE with corresponding local conserved charges. However, it is in general necessary also to include parafermionic charges. Finally, we showed that irrelevant perturbations of the CFT hamiltonian involving powers of the stress tensor lead to dependence of the propagation speed on the initial state, and to a diffusive-like broadening of the horizon. Although we have treated these three topics independently, they could clearly be combined, at some technical expense, into a unified picture, considering, for example, the convergence of the reduced density matrix to that of a GGE in the case of a general initial state, in a perturbed CFT. 

There has been some discussion in the literature as to exactly which conserved charges should be taken into account in the GGE \cite{discussion}. The results from lattice studies initially suggested that only `local' operators should be included, defined as those which involve lattice spins over a finite range of lattice spacings. However these have been found to be incomplete in some cases, and this appears to be associated with the existence of bound states \cite{bound}. 
However, recently this problem has been repaired for the case of Heisenberg chains by the addition of `quasi-local' charges \cite{quasilocal}.
It has been similarly argued within integrable field theory that the usual local conserved charges are not sufficient to describe the GGE \cite{EsslerMuss}.

The case of a rational CFT, however, appears to be different: in a finite but large system the eigenstates of the hamiltonian are in correspondence with the states of a finite number of Virasoro modules (or those of an extended symmetry) and these states are in bijection with the Virasoro conserved charges (i.e.~those made up of the stress tensor and its derivatives.) Thus, giving the expectation value of the conserved charges in any initial state should specify that state uniquely. (In fact, because of the existence of null states, the Virasoro charges are over-complete.) This argument suggests that for a rational CFT,  the non-Abelian GGE is sufficient. One might ask whether less information is needed to specify the reduced density matrix $\rho_\ell$ in a subsystem of length $\ell\ll L$. In fact, by considering propagation along the cylinder, it can be seen that $\rho_\ell$ is determined in this limit by only the ground state of the hamiltonian $h+\delta h$ in (\ref{deltah}), not its full spectrum. Since this hamiltonian is defined on a system of length $\beta\ll\ell$, one might argue that this requires less information that specifying the initial state. However, this argument compares two infinities and it seems difficult to make more precise.

Nevertheless the values of the charges in the commuting sub-algebra given in \cite{Zam}, are insufficient to specify $\rho_\ell$, since it retains perfect memory of all the left- and right-movers in the initial state in regions of length $\ell$. This feature appears to be related to the exactly linear dispersion relation in a CFT. A similar result has recently been found by Sotiriadis \cite{Sot} for the case of a free boson evolving from a non-gaussian initial state. Although this is a CFT, it is peculiar in possessing U$(1)$ conserved currents $(J,\bar J)$, out of which local conserved quantities may be formed by taking products and derivatives. (Alternatively, the conserved quantities are the (non-local) number operators of the momentum modes of the field. The linear dispersion relation appears to play a crucial role for this model, since it was shown \cite{CS} that for a massive free boson memory of the initial state is lost and the number-operator GGE is sufficient.

However, our argument does not necessarily extend to CFTs where the central charge is greater than the number of primary conserved currents, with an infinity of primary fields.   These CFTs are believed to be non-integrable, or chaotic, in a well-defined sense \cite{Hart}. Nor does it apply to massive integrable field theories viewed as perturbed CFTs, since in this case only some of the Virasoro charges remain conserved.

The parafermionic operators we claim that one must also introduce are better described in CFT language as `semi-local': they depend on single space-time coordinate but depend non-locally on the state of the system described in the basis of local operators. Nevertheless they appear in the spectrum of the CFT, for example on a cylinder, if suitable boundary conditions are imposed, and in some cases, such as the Ising fermion, are more fundamental than the local fields of the theory in the sense that the latter may be built out of them. It is interesting that it should be in principle possible to access their correlation functions in the quench scenario according to the above arguments.

It would be interesting to extend our analysis of the non-integrable $T\overline T$ perturbation of the CFT to other situations, for example an inhomogeneous quench \cite{CSot}, where there are non-zero energy currents, or to the steady state currents set up by a non-zero temperature gradient \cite{BD}. Recently this problem has been addressed from a different point of view \cite{BD2}.

\ack
Some parts of this paper have taken a long time to see the light of day. The work on GGEs was first announced at a workshop at the Galileo Galilei Institute in Florence in 2012 and an improved version at a conference at CUNY in 2013. Additional work was done during programs of the Kavli Institute for Theoretical Physics in Santa Barbara in 2014 and 2015. Meanwhile a paper has appeared by Mandal, Sinha and Sorokhaibam \cite{MSS} in which it is shown by very similar methods that the CC initial state, perturbed by additional conserved charges, leads to the GGE ansatz for the stationary reduced density matrix of a subsystem. I am grateful to G.~Mandal for pointing out this reference.

This research was supported in part by the National Science Foundation under Grant No.~NSF PHY11-25915, and a grant from the Simons Foundation. The author particularly thanks Fabian Essler for many instructive conversations about quantum quenches in general and about GGEs in particular, and Pasquale Calabrese for a reading of the first version of this paper. He also thanks D.~Bernard, T.~Hartman, M.~Fagotti,  J.~March-Russell and S.~Sotiriadis for discussions and correspondence, and the referees for several important comments on the first version of this paper.

\end{document}